\documentclass[prd,showpacs,preprintnumbers,amsmath,amssymb]{revtex4}

\usepackage{graphicx}
\usepackage{dcolumn}
\usepackage{bm}

\begin{document}

\title{Improved Approximations for Fermion Pair Production \\
in Inhomogeneous Electric Fields}

\author{Sang Pyo Kim}\email{sangkim@kunsan.ac.kr}

\affiliation{Department of Physics, Kunsan National University,
Kunsan 573-701, Korea\footnote{Permanent address},}
\affiliation{Asia Pacific Center for Theoretical Physics, Pohang
790-784, Korea}

\author{Don N. Page}\email{don@phys.ualberta.ca}

\affiliation{Theoretical Physics Institute, Department of Physics,
University of Alberta, Edmonton, Alberta, Canada T6G 2G7}

\date{2007 January 9}

\begin{abstract}Reformulating the instantons in a complex plane for
tunneling or transmitting states, we calculate the pair-production
rate of charged fermions in a spatially localized electric field,
illustrated by the Sauter electric field $E_0~{\rm sech}^2 (z/L)$,
and in a temporally localized electric field such as $E_0~{\rm
sech}^2 (t/T)$. The integration of the quadratic part of WKB
instanton actions over the frequency and transverse momentum leads
to the pair-production rate obtained by the worldline instanton
method, including the prefactor, of Phys. Rev. D {\bf 72}, 105004
(2005) and {\bf 73}, 065028 (2006). It is further shown that the
WKB instanton action plus the next-to-leading order contribution in
spinor QED equals the WKB instanton action in scalar QED, thus
justifying why the WKB instanton in scalar QED can work for the
pair production of fermions.  Finally we obtain the pair-production
rate in a spatially localized electric field together with a
constant magnetic field in the same direction.

\end{abstract}

\pacs{12.20.-m, 13.40.-f \hfill  Alberta-Thy-01-07}

\maketitle

\section{Introduction}

The physics of strong electromagnetic fields, in particular vacuum
polarization and pair production, has been studied by Sauter
\cite{Sauter31}, Heisenberg and Euler \cite{Heisenberg-Euler}, and
Weisskopf \cite{Weisskopf} even before the advent of quantum
electrodynamics (QED). Using the proper-time path integral,
Schwinger obtained the one-loop QED effective action in a constant
electromagnetic field \cite{Schwinger}. The imaginary part of the
effective action under the influence of electric fields leads to
the decay rate of the vacuum and thereby the pair-production rate
of charged particles. (For a recent review and references on QED
and pair production, see \cite{Dunne}.) One approach to pair
production is tunneling of virtual pairs from the Dirac sea
\cite{Keldysh,Nikishov-Ritus,Brezin-Itzykson,
Narozhnyi-Nikishov,Popov,CNN,Kim-Page,Kim-Page2}, in which
instantons determine the tunneling probability and thereby the
pair-production rate. Another approach is the recent worldline
method \cite{Schubert,Gies-Langfeld,LMG,
MLG,Bast-Schubert,Gies-Klingmuller,Dunne-Schubert,DWGS,Dunne-Wang}.

On the other hand, physics of strong fields in laboratory have
recently attracted much attention due to the rapid development of
laser technology. In the near future the X-ray free electron lasers
from Linac Coherent Light Source at SLAC \cite{slac} and TESLA at
DESY \cite{desy} are expected to produce at the focus an electric
field near the critical strength $E_c = m^2c^3/e \hbar ~(1.3 \times
10^{18}~{\rm V/m})$ for electron-position pair production. Also
ultrahigh intense lasers have been developed using a technique of
amplifying pulses of picosecond and few femtosecond time scales (for
a review and references, see \cite{Mourou}). The focal region of
colliding lasers may correspond to a strong QED regime, in which not
only vacuum polarization but also pair production can be tested
\cite{Ringwald} (see also \cite{Marklund} for other QED related
physics). The localized beam in space and time does necessarily
imply inhomogeneous electromagnetic fields. Thus pair production of
charged particles by inhomogeneous electric fields is not only a
theoretical issue but also an experimental concern.

In previous papers \cite{Kim-Page,Kim-Page2}, we formulated the
pair-production rate of charged particles in inhomogeneous electric
electric fields in terms of the instantons for tunneling states in
the space-dependent (Coulomb) gauge, and worked out explicitly the
rate for the Sauter electric field \cite{Sauter32} in the WKB
approximation. Recently, using the worldline path integral, Dunne
and Schubert \cite{Dunne-Schubert} obtained worldline instantons in
a gauge-independent way, and Dunne, Wang, Gies and Schubert
\cite{DWGS} further found the prefactor to the worldline instantons
to calculate the pair-production rate by inhomogeneous fields.

In this paper, using the phase-integral method
\cite{Froman,Dunham}, we further elaborate the instanton method of
Refs. \cite{Kim-Page,Kim-Page2} by defining the instanton action in
a gauge-independent way as a contour integral in the complex space
or time plane. The divergence problem at turning points in the WKB
method can be naturally avoided by the contour integral excluding
the branch cut connecting two complex roots corresponding to
turning points. We use the new instanton method to calculate the
leading-order (LO) and next-to-leading-order (NLO) WKB instanton
actions both in a spatially localized electric field $E_0~{\rm
sech}^2 (z/L)$ and in a temporally localized electric field
$E_0~{\rm sech}^2 (t/T)$ of Sauter type \cite{Sauter32}, and then
obtain the pair-production rate for charged fermions. We then
compare the pair-production rate with that obtained by the
worldline instanton method \cite{Dunne-Schubert,DWGS} and with the
exact result by Nikishov \cite{Nikishov}.

We show that for the Sauter electric field \cite{Sauter32}, the
minimum value of our WKB instanton action ${\cal S}^{(0)}_{{\bf
k}_{\perp}}(\omega) = 2S^{(0)}_{{\bf k}_{\perp}}(\omega)$, as a
function of the transverse momentum ${\bf k}_{\perp}$ and the
frequency $\omega$, is the same as that of the single-worldline
instanton action of Ref. \cite{Dunne-Schubert}, and that the
gaussian approximation for the integrals over $\omega$ and ${\bf
k}_{\perp}$ in our approach using ${\cal S}^{(0)}_{{\bf
k}_{\perp}}(\omega)$ gives the same pair-production rate, including
the prefactor \cite{DWGS}, as the worldline instanton method. We
further calculate the NLO WKB actions ${\cal S}^{(2)}_{sp, {\bf
k}_{\perp}}(\omega)$ in spinor QED and show that the sum of the WKB
and NLO instanton actions in spinor QED is equal to the WKB
instanton action in scalar QED.  Finally we study pair production
in the Sauter electric field together with a constant magnetic
field in the same direction.

The organization of this paper is as follows. In Sec. II, we
reformulate the instanton method in a gauge-independent way in the
complex plane of space or time. We also show that the terms in the
WKB approximation for a constant electric field vanish beyond the
leading order, and the WKB approximation thus reproduces the
well-known exact one-loop result from the point of view of the
instanton method. In Sec. III, we calculate within the
leading-order WKB approximation the instanton actions in scalar QED
for a localized electric field of spatial or temporal Sauter-type
and obtain the pair-production rate for both cases. We find that
the WKB instanton action in scalar QED gives a pair-production
rate closer to the exact result for spinor QED up to higher order
in an adiabaticity parameter. In Sec. IV, we calculate the WKB and
NLO instanton actions in spinor QED and find that the instanton
action up through the NLO contribution for each frequency and
transverse momentum equals the WKB instanton action in scalar
QED, thus justifying the reason why the WKB instanton action in
scalar QED gives a good result. In Sec. V, we find the WKB
instanton action in the Sauter electric field together with a
constant magnetic field in the same direction.

\section{Gauge Independence of Instanton Actions}

The worldline instanton method is based on Feynman's worldline path
integral \cite{Feynman}, manifestly a gauge invariant formalism,
and calculates the instanton action from a closed instanton
trajectory \cite{Gies-Langfeld,LMG,
MLG,Bast-Schubert,Gies-Klingmuller,Dunne-Schubert,DWGS,Dunne-Wang}.
Affleck, Alvarez and Manton found the instanton action in a
constant electric field from the closed instanton trajectory
\cite{Affleck}. In this paper we show how the instanton method of
Refs. \cite{Kim-Page,Kim-Page2} can be reformulated by the same
contour integral for two different gauges. For the sake of
simplicity we first focus on the constant electric field, which can
be treated exactly, and then extend to inhomogeneous fields in the
next section. The electric field can be written in terms of two
simple alternative gauges: the space-dependent (Coulomb) gauge and
the time-dependent gauge. We shall first work on scalar QED,
described by the Klein-Gordon equation, and then discuss spinor
QED, described by the Dirac equation, in Sec. IV. The Klein-Gordon
equation for charged particles with charge $q$ ($q > 0$) and mass
$m$ takes the form [in units with $\hbar = c = 1$ and with metric
signature $(+, -, -, -)$]
\begin{equation}
 [\eta^{\mu \nu} (\partial_{\mu} + i q A_{\mu}) (\partial_{\nu}
   + i q A_{\nu}) + m^2 ] \Phi = 0.
\label{kg eq}
\end{equation}

In the space-dependent gauge, the Klein-Gordon equation, after
being decomposed into Fourier modes, becomes a tunneling problem,
the so-called sub-barrier penetration. The essence of the instanton
method is that the tunneling states of this equation lead to the
vacuum decay and pair production of charged particles. On the other
hand, in the time-dependent gauge, the mode-decomposed equation
resembles a one-dimensional scattering problem over a potential
barrier, the so-called super-barrier transmission. The
positive-frequency modes define the vacuum state at past infinity,
and the negative-frequency modes at future infinity lead to the
number of created pairs \cite{Parker,DeWitt}. Fr\"{o}man and
Fr\"{o}man \cite{Froman} showed that the probability amplitudes of
both the tunneling and the transmitted states could be found in
terms of a contour integral in the complex plane of space or time.
This contour integral yields the same WKB instanton action in both
gauges for a constant electric field.

\subsection{Constant E-field in the space-dependent gauge}

In the space-dependent (Coulomb) gauge, a constant electric field
along the $z$-direction has the potential $A_{\mu} = (-Ez, 0, 0,
0)$. The mode-decomposed Klein-Gordon equation in (\ref{kg eq}) now
takes the form
\begin{eqnarray}
[ - \partial_z^2 + Q(z) ] \phi_{\omega {{\bf k}_{\perp}}} (z) = 0,
\end{eqnarray}
where
\begin{eqnarray}
Q(z) = m^2 + {\bf k}_{\perp}^2 - (\omega + qEz)^2. \label{q1}
\end{eqnarray}
Each mode equation now describes the tunneling problem under the
upside-down harmonic potential barrier and has two real turning
points
\begin{eqnarray}
z_{\pm} = - \frac{\omega}{qE} \pm \sqrt{\frac{m^2 + {\bf
k}_{\perp}^2}{(qE)^2}}.
\end{eqnarray}
The exact wave function is given by the complex parabolic
cylindrical function $E(S_{{\bf k}_{\perp}}/\pi, \sqrt{2/qE} (q zE
+\omega))$ in terms of the instanton action $S_{{\bf k}_{\perp}} =
\pi(m^2 + {\bf k}^2_{\perp})/ (2 qE)$ \cite{Kim-Page}.

Using the phase-integral formula \cite{Froman,Dunham}, the wave
function can be written in one asymptotic region $z \ll z_-$ as
\begin{eqnarray}
\phi_{\omega {\bf k}_{\perp}} (z) = A \Bigl(e^{\frac{i \pi}{4}}
\varphi_{\omega {\bf k}_{\perp}} (z) \Bigr)+  B \Bigl( e^{ \frac{i
\pi}{4}} \varphi_{\omega {\bf k}_{\perp}} (z) \Bigr)^*,
\end{eqnarray}
where
\begin{equation}
A = (e^{2 S_{\bf k}} +1 )^{1/2}, \quad B = e^{S_{\bf k}},
\end{equation}
and $\varphi_{\omega {\bf k}_{\perp}} (z)$ has unit incoming flux
(with the group velocity away from the barrier on the back
side). Here the leading-order WKB instanton action is given by
\begin{eqnarray}
S^{(0)}_{{\bf k}_{\perp}} = \int_{z_-}^{z_+} \sqrt{Q(z)} dz =
\frac{\pi(m^2 + {\bf k}_{\perp}^2)}{2qE},
\end{eqnarray}
which is the exact action for the constant $E$-field, that is, $
S_{{\bf k}_{\perp}} = S^{(0)}_{{\bf k}_{\perp}}$. In the other
asymptotic region $z \gg z_+$,  the tunneling wave function is given
by
\begin{eqnarray}
\phi_{\omega {\bf k}_{\perp}} (z) = \Bigl( e^{- \frac{i \pi}{4}}
\varphi_{\omega {\bf k}_{\perp}} (z) \Bigr)^*.
\end{eqnarray}

Now the action can also be defined in the complex $z$ plane by the
contour integral \cite{Dunham,Froman}
\begin{eqnarray}
{\cal S}_{{\bf k}_{\perp}} =  2 S_{{\bf k}_{\perp}} = - i
\oint_{\Gamma_{K}} q(z) dz, \label{cont int}
\end{eqnarray}
where the integral is along the contour in  Fig. 1. Here,
\begin{eqnarray}
q(z) = \sqrt{-Q(z)} \sum_{n = 0}^{\infty} Y_{2n}(z),
\end{eqnarray}
and the first two $Y_{2n}$ are \cite{Froman,Dunham}
\begin{eqnarray}
Y_0 (z) &=& 1, \nonumber\\
Y_2 (z) &=& - \frac{1}{32 Q^3} \Biggl[5 \Bigl(\frac{dQ}{dz}
\Bigr)^2 - 4 Q \frac{d^2 Q}{dz^2} \Biggr].
\end{eqnarray}
Note that the complex function $\sqrt{- Q}$ has a branch cut along
the real line segment connecting two roots $z_{\pm}$ and is
single-valued outside the closed loop of Fig. 1 \cite{Arfken}. From
the Laurent expansion for large $z$,
\begin{eqnarray}
\sqrt{-Q} = qE \Biggl[z - \frac{z_+ + z_-}{2} - \frac{(z_+ -
z_-)^2}{8 z} - \cdots \Biggr],
\end{eqnarray}
we find the residue $qE (z_+ - z_-)^2/8$ at the simple pole at $z =
\infty$ \cite{Markushevich}. Hence the contour integral in the
exterior region of the closed loop in Fig. 1 leads to the
leading-order WKB instanton action
\begin{eqnarray}
{\cal S}^{(0)}_{{\bf k}_{\perp}} = 2 \pi i \Bigl[ - i qE \frac{(z_+
- z_-)^2}{8} \Bigr] =  \frac{\pi(m^2 + {\bf k}_{\perp}^2)}{qE}.
\end{eqnarray}
The WKB instanton action agrees with Eq. (13) of Ref.
\cite{Kim-Page} and Eq. (21) of Ref. \cite{Kim-Page2}, and also
with Eq. (27) of Ref. \cite{Dunne-Schubert} for the action of the
worldline instanton.

\begin{figure}[t]
{\includegraphics[width=0.6\linewidth,height=0.05\textheight
]{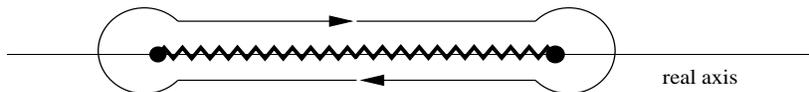}} \caption{The contour integral in the complex plane
$z$ excluding a branch cut connecting two real roots $z_{\pm}$.}
\end{figure}

\subsection{Constant E-field in the time-dependent gauge}

In the time-dependent gauge, the potential is given by $A_{\mu} =
(0, 0, 0, -Et)$. The mode-decomposed Klein-Gordon equation of
(\ref{kg eq}) takes the form
\begin{eqnarray}
[ - \partial_t^2 + Q_t(t) ] \phi_{\omega {\bf k}} (t) = 0,
\end{eqnarray}
where
\begin{eqnarray}
Q_t(t) = - \Bigl[ m^2 + {\bf k}_{\perp}^2 + (k_z - qEt)^2 \Bigr].
\label{q2}
\end{eqnarray}
The problem now becomes the super-barrier transmission over the
upside-down harmonic potential. The vacuum state is defined in terms
of the positive-frequency (adiabatic) solution at past infinity $(t
= - \infty)$. Particle production is ascribed to the negative
frequency (adiabatic) solution at future infinity $(t = \infty)$.

The function $\sqrt{-Q_t}$ now has a branch cut along a line segment
parallel to the imaginary axis in the complex $t$ plane, which
connects two complex roots
\begin{eqnarray}
t_{\pm} =  \frac{k_z}{qE} \pm i \sqrt{\frac{m^2 + {\bf
k}_{\perp}^2}{(qE)^2}}.
\end{eqnarray}
Then the transmission probability is determined by the contour
integral \cite{Froman}
\begin{eqnarray}
{\cal S}_{{\bf k}_{\perp}}  = 2 S_{{\bf k}_{\perp}} = i
\oint_{\Gamma_{K}} \sqrt{- Q_t (t)} dt, \label{ti int}
\end{eqnarray}
along the contour in Fig. 2. As $\sqrt{- Q_t}$ has a simple pole at
$t = \infty$,
\begin{eqnarray}
\sqrt{-Q_t} = qE \Biggl[t - \frac{t_+ + t_-}{2} - \frac{(t_+ -
t_-)^2}{8 t} - \cdots \Biggr],
\end{eqnarray}
and the residue is $qE (t_+ - t_-)^2/8$, we obtain the WKB action
\begin{eqnarray}
{\cal S}_{{\bf k}_{\perp}} = 2 \pi i \Bigl[ i qE \frac{(t_+ -
t_-)^2}{8} \Bigr] = \frac{\pi(m^2 + {\bf k}_{\perp}^2)}{qE}.
\end{eqnarray}

Though the Klein-Gordon equation in the time-dependent gauge
potential for the constant $E$-field is equivalent to the scattering
problem over a potential barrier, the transmission probability is
determined by the action now defined in the complex time plane. This
holds true for general time-dependent electric fields. In this
sense, the action defined by the contour integral (\ref{cont int})
or (\ref{ti int}) in the complex plane of space or time is the same
for both gauge potentials.

\begin{figure}[t]
{\includegraphics[width=0.15\linewidth,height=0.25\textheight
]{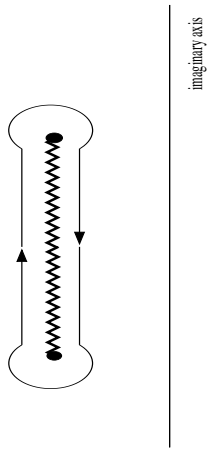}} \caption{The contour integral in the complex plane
$t$ excluding a branch cut connecting two complex roots $z_{\pm}$.}
\end{figure}

\subsection{$S^{(2)} = 0$ for constant E-field}

In the space-dependent gauge, the next-to-leading order (NLO) WKB
contribution to the instanton action takes the form
\begin{eqnarray}
{\cal S}^{(2)}_{{\bf k}_{\perp}} = i \oint_{\Gamma_{K}} \sqrt{- Q
(z)} \, \frac{1}{32 Q^3} \Biggl[5 \Bigl(\frac{dQ}{dz}
\Bigr)^2 - 4 Q \frac{d^2 Q}{dz^2} \Biggr] dz \label{c-int2}
\end{eqnarray}
for the same contour in Fig. 1. The simple roots $z_{\pm}$ of $Q(z)$
cannot give simple poles, as they are excluded by the contour.
Further the integrand does not have a simple pole at infinity, since
it has the expansion of the form
\begin{eqnarray}
\frac{1}{32 Q^3} \Biggl[5 \Bigl(\frac{dQ}{dz} \Bigr)^2 - 4 Q
\frac{d^2 Q}{dz^2} \Biggr] = \sum^{\infty}_{n = 4}
\frac{a_{n}}{z^n}.
\end{eqnarray}
Therefore, the contour integral (\ref{c-int2}) vanishes, and ${\cal
S}^{(2)}_{{\bf k}_{\perp}} = 0$ for the constant $E$-field. As all
higher order actions vanish by the same argument, the WKB instanton
actions agree with the result from the exact solution of the
Klein-Gordon equation. Also the same argument holds for the
time-dependent gauge. The integral over all the transverse momentum
recovers the well-known pair-production rate in a constant electric
field \cite{Schwinger}.

\section{Localized Electric Fields}

Consider a static plane-symmetric but $z$-dependent electric field
$E(z)$ in the $z$-direction of maximum value $E_0$ and of effective
length $L$ defined so that
\begin{eqnarray}
E_0 L \equiv \frac{1}{2} \int_{-\infty}^{\infty} E(z) dz.
\end{eqnarray}
It may be characterized by the parameter
\begin{eqnarray}
\epsilon \equiv \frac{m}{q E_0 L}.
\end{eqnarray}

Similarly, a homogeneous time-dependent electric field $E(t)$ in the
$z$-direction of maximum value $E_0$ and of effective time $T$
defined so that
\begin{eqnarray}
E_0 T \equiv \frac{1}{2} \int_{-\infty}^{\infty} E(t) dt
\end{eqnarray}
may be characterized by the parameter
\begin{eqnarray}
\epsilon_t \equiv \frac{m}{q E_0 T}.
\end{eqnarray}

Pair production is energetically allowed for $\epsilon  < 1$ and for
any $\epsilon_t$, though it is strongly suppressed for $\epsilon_t
\gg 1$ . Pair production by localized electric fields significantly
differs from that by the constant electric field due to a size
effect \cite{Wang-Wong}. The finite size effect on pair production
was also shown in many inhomogeneous electric fields
\cite{Kim-Page,Dunne-Schubert,Kim-Page2,DWGS}.

We shall also define another parameter,
\begin{eqnarray}
\delta \equiv \frac{qE_0}{\pi m^2},
\end{eqnarray}
which (for $\epsilon < 1$ or $\epsilon_t < 1$) is small when the
pair-production rate is small. A third useful parameter is the
following combination of the previous two:
\begin{eqnarray}
b \equiv \frac{\delta}{\sqrt{1-\epsilon^2}} \quad \mathrm{or} \quad
b_t \equiv \frac{\delta}{\sqrt{1 + \epsilon_t^2}}. \label{bb}
\end{eqnarray}
When the WKB approximation is good, then $b \ll 1$ or $b_t \ll 1$, so $b$ or
$b_t$ serves as an adiabaticity parameter.

In this section we consider the spatially localized electric field
$E(z) = E_0 ~{\rm sech}^2 (z/L)$ in the $z$-direction given by the
Sauter potential \cite{Sauter32}
\begin{equation}
A_0 (z) = - E_0 L \tanh \Bigl(\frac{z}{L} \Bigr).
\end{equation}
From now on $E(z) = E_0 ~{\rm sech}^2 (z/L)$ will be called the
Sauter electric field \cite{Sauter32}. Pairs are produced only when
$ q E_0 L - m \geq \omega \geq -(q E_0 L -m)$. This electric field
varies slowly over the effective length $L$. The other field is a
uniform but temporally localized electric field $E (t) = E_0 ~{\rm
sech}^2 (t/T)$, effectively lasting over the period $T$.

\subsection{Sauter Electric Field}

In the first case of the Sauter potential $A_0 (z) = - E_0 L \tanh
(z/L)$ \cite{Sauter32}, we change the variable $\zeta = L \tanh
(z/L)$ to write the leading-order WKB action as
\begin{eqnarray}
{\cal S}^{(0)}_{{\bf k}_{\perp}} = 2 S_{{\bf k}_{\perp}} = - i
\oint_{\Gamma_{K}} \frac{\sqrt{-Q(\zeta)}}{1 - \frac{\zeta^2}{L^2}}
d\zeta, \label{sc int}
\end{eqnarray}
where
\begin{eqnarray}
Q(\zeta) = m^2 + {\bf k}_{\perp}^2 - (\omega + q E_0 \zeta)^2.
\end{eqnarray}
As $|\zeta| \leq L$, we expand the integrand as a Laurent series
\begin{eqnarray}
\frac{\sqrt{-Q(\zeta)}}{1 - \frac{\zeta^2}{L^2}} = q E_0 \sum_{n =
0}^{\infty} \sum_{l = 0}^{\infty} \frac{C_l}{L^{2n}} \zeta^{2n +1 -
l},
\end{eqnarray}
where we also expand the square root for large $\zeta$ as
\begin{eqnarray}
f^{(0)}(\zeta) \equiv \sqrt{-\frac{Q(\zeta)}{(qE_0 \zeta)^2}}  =
\sum^{\infty}_{l = 0} \frac{C_{l}}{\zeta^l}.
\end{eqnarray}
The integrand of (\ref{sc int}) has simple poles at large $\zeta$
for $l = 2n + 2 ~(\geq 2$), and the sum of residues is
\begin{eqnarray}
\sum_{residue} = - qE_0 L^2 \sum_{n = 1}^{\infty}
\frac{C_{2n}}{L^{2n}}.
\end{eqnarray}
Noting that
\begin{eqnarray}
f^{(0)}(L) + f^{(0)} (-L) = 2 \sum_{n = 0}^{\infty}
\frac{C_{2n}}{L^{2n}},
\end{eqnarray}
we finally obtain the leading-order WKB action
\begin{eqnarray}
{\cal S}^{(0)}_{{\bf k}_{\perp}} = \pi qE_0 L^2 \Bigl[ 2 -
f^{(0)}(L) - f^{(0)} (-L) \Bigr].
\end{eqnarray}
Therefore, the leading-order WKB action in scalar QED can be written
as
\begin{eqnarray}
{\cal S}^{(0)}_{{\bf k}_{\perp}} = \frac{Z}{2} \Bigl[2 - \sqrt{(1+
\mu)^2 - \epsilon^2 (1 + \kappa^2) } - \sqrt{(1 - \mu)^2 -
\epsilon^2 (1 + \kappa^2) }\Bigr], \label{sc ins}
\end{eqnarray}
in terms of the dimensionless scaled variables
\begin{eqnarray}
\mu \equiv \frac{\omega}{qEL}, \quad \kappa \equiv \frac{k_{\perp}}{m},
\end{eqnarray}
and the dimensionless parameters
\begin{eqnarray}
Z = 2 \pi q E_0 L^2 = \frac{2}{\delta \epsilon^2}, \quad \epsilon =
\frac{m}{q E_0 L}, \quad \delta = \frac{qE_0}{\pi m^2}.
\end{eqnarray}

We shall first compare the scalar QED instanton actions with other
results. As ${\cal S}^{(0)}_{{\bf k}_{\perp}}$ is an even function
of $\mu$ and $\kappa$, we may expand the action (\ref{sc ins})
as a power series in $\mu$ and $\kappa$,
\begin{eqnarray}
{\cal S}^{(0)}_{{\bf k}_{\perp}} = {\cal S}^{(0)}_{\{0\}} + {\cal
S}^{(0)}_{\{2\}} + {\cal S}^{(0)}_{\{4\}} + \sum_{n = 3}^{\infty}
{\cal S}^{(0)}_{\{ 2n \}},
\end{eqnarray}
where the first few ${\cal S}^{(0)}_{\{2n\}}$ are
\begin{eqnarray}
{\cal S}^{(0)}_{\{0\}} &=& Z (1 - \sqrt{1 - \epsilon^2}), \nonumber\\
{\cal S}^{(0)}_{\{2\}} &=& \frac{ \mu^2}{\delta (1 -
\epsilon^2)^{3/2}} + \frac{\kappa^2}{\delta (1 -
\epsilon^2)^{1/2}}, \nonumber\\
{\cal S}^{(0)}_{\{4\}} &=& \frac{1}{4 \delta (1 - \epsilon^2)^{7/2}}
[(4 + \epsilon^2) \mu^4  + 2(1 - \epsilon^2) (2 + \epsilon^2) \mu^2
\kappa^2 + \epsilon^2 (1 - \epsilon^2)^2 \kappa^4 ].
\end{eqnarray}
In the special case of $\omega = 0$ and ${\bf k}_{\perp} = 0$,
we obtain the action
\begin{eqnarray}
{\cal S}^{(0)}_{\{0\}}
 = \frac{1}{\delta} \frac{2}{1 + \sqrt{1 - \epsilon^2}}
 = Z (1 - \sqrt{1 - \epsilon^2})
 = \frac{2}{\sqrt{1 - \epsilon^2} + 1 - \epsilon^2}\frac{1}{b}
 > \frac{1}{b} \gg 1,
\end{eqnarray}
agreeing with the action of the worldline instanton (62) of Ref.
\cite{Dunne-Schubert}. This is also true for spinor QED as will be
shown in Sec. IV.

Weighting the gaussian integral of $e^{-({\cal S}^{(0)}_{\{0\}} +
{\cal S}^{(0)}_{\{2\}})}$ over $\omega$ and ${\bf k}_{\perp}$ by a
power series expansion of $e^{-\sum_{n =2} {\cal
S}^{(0)}_{\{2n\}}}$ through 8th order in $\omega$ and ${\bf
k}_{\perp}$, we obtain a WKB approximation for the pair-production
rate of charged particles per unit time and unit cross-sectional
area,
\begin{eqnarray}
{\cal N}^{(0)} &=& \frac{2}{(2 \pi)^3} \int d \omega \int d^2 {\bf
k}_{\perp} e^{- {\cal S}^{(0)}_{\bf k}} \nonumber\\
& \approx & \frac{(qE_0)^{5/2}L}{4 \pi^3 m} (1 - \epsilon^2)^{5/4}
e^{- Z (1 - \sqrt{1 - \epsilon^2})} \Biggl[1 - \frac{5}{16}
(4+3\epsilon^2)b + \frac{105}{512}(4-\epsilon^2)^2 b^2
\nonumber\\&& -\frac{315}{8192}
(320-432\epsilon^2+124\epsilon^4-\epsilon^6) b^3 \Biggr].
\label{WKB00}
\end{eqnarray}
Here and hereafter we insert a factor of 2, for two charged scalar
fields, to compare with the results of spinor QED with two spins
for each spin-$1/2$ fermion state.  Thus all single-scalar QED
results would be half of those listed here.

The contribution up to the quadratic terms of the WKB instanton
action, the factor in front of the square bracket, agrees with Eq.
(4.7) of Ref. \cite{DWGS} from the worldline instanton
approximation. The 2nd, 3rd, and 4th terms in the square bracket
are the contributions from the 4th, 6th, and 8th order terms of the
WKB instanton action, though these terms in the action are
intertwined in the pair-production rate because of the
exponentiation of the action.

It is interesting to compare this leading-order WKB result in scalar
QED with the exact result in spinor QED \cite{Nikishov}, given in
terms of $Z$ and $\epsilon$ by the double integral in Eq. (62) of
\cite{Kim-Page2}.  For $b \ll 1$, so that one may drop terms that
are exponentially smaller than the dominant terms by factors like
$e^{-{\cal S}^{(0)}_{\{0\}}} < e^{-1/b}$, the leading approximation
to the double integral, given by Eq. (64) of \cite{Kim-Page2},
agrees with the leading term of Eq. (\ref{WKB00}), the coefficient
in front of the square bracket.  We have checked that taking
additional dominant terms of the double integral (suppressed not
exponentially but only by powers of $\mu$ and $\kappa$) indeed gives
precisely the same total result as the extreme right hand side of
Eq. (\ref{WKB00}).  Indeed, one can show that the entire
leading-order WKB approximation for scalar QED pair production in
the Sauter potential, given by the first right hand side of Eq.
(\ref{WKB00}), immediately after the $=$ sign, is equal to the total
power series of all dominant terms of the exact result for spinor
QED when one drops only terms that are exponentially suppressed.

\subsection{Temporally Localized Field}

In the second case of the temporally localized electric field, we
change the variable $\tau = T \tanh (t/T)$ to write the WKB actions
as
\begin{eqnarray}
{\cal S}^{(0)}_{{\bf k}_{\perp}} = 2 S_{{\bf k}_{\perp}} =  i
\oint_{\Gamma_{K}} \frac{\sqrt{-Q_t(\tau)}}{1 -
\frac{\tau^2}{T^2}} d\tau, \label{tint}
\end{eqnarray}
where
\begin{eqnarray}
Q_t(\tau) = - \Bigl[ m^2 + {\bf k}_{\perp}^2 + (k_z + q E_0 \tau)^2
\Bigr].
\end{eqnarray}
Here and hereafter the subscript $t$ denotes any quantity in the
time-dependent gauge. The integral (\ref{tint}) can be obtained from
Eq. (\ref{sc int}) for the spatially localized electric field by
replacing $L$ by $T$ and $Q$ from Eq. (\ref{q1}) by $Q_t$ from Eq.
(\ref{q2}). In terms of the scaled variables
\begin{eqnarray}
\mu_t = \frac{k_z}{qE_0T}, \quad \kappa = \frac{k_{\perp}}{m},
\end{eqnarray}
and parameters
\begin{eqnarray}
Z_t = 2 \pi qE_0 T^2 = \frac{2}{\delta \epsilon_t^2}, \quad
\epsilon_t = \frac{m}{qE_0T}, \quad \delta = \frac{qE_0}{\pi m^2},
\end{eqnarray}
we have
\begin{eqnarray}
{\cal S}^{(0)}_{{\bf k}_{\perp}} &=& \frac{Z_t}{2} \Bigl[ \sqrt{(1+
\mu_t)^2 + \epsilon_t^2 (1 + \kappa^2) } + \sqrt{(1 - \mu_t)^2 +
\epsilon_t^2 (1 + \kappa^2) } - 2\Bigr].
\end{eqnarray}
In the special case of $k_z = 0$ and ${\bf k}_{\perp} = 0$, the
action is
\begin{eqnarray}
{\cal S}^{(0)}_{\{0\}}
 = \frac{1}{\delta} \frac{2}{1 + \sqrt{1 + \epsilon_t^2}}
 = Z_t (\sqrt{1 + \epsilon_t^2} -1)
 = \frac{2}{\sqrt{1 + \epsilon_t^2} + 1 + \epsilon_t^2}\frac{1}{b_t}
 < \frac{1}{b_t},
\end{eqnarray}
which agrees with Eq. (38) of Ref. \cite{Dunne-Schubert}.

As in the case of the Sauter electric field, we expand the
instanton action in power of $\mu_t$ and $\kappa$ as
\begin{eqnarray}
{\cal S}^{(0)}_{{\bf k}_{\perp}} = {\cal S}^{(0)}_{\{0\}} + {\cal
S}^{(0)}_{\{2\}} + {\cal S}^{(0)}_{\{4\}} + \sum_{n = 3}^{\infty}
{\cal S}^{(0)}_{\{2n \}},
\end{eqnarray}
where
\begin{eqnarray}
{\cal S}^{(0)}_{\{0\}} &=& Z_t (\sqrt{1 + \epsilon_t^2} -1), \nonumber\\
{\cal S}^{(0)}_{\{2\}} &=& \frac{ \mu_t^2}{\delta (1 +
\epsilon_t^2)^{3/2}} + \frac{\kappa^2}{\delta (1 +
\epsilon_t^2)^{1/2}}, \nonumber\\
{\cal S}^{(0)}_{\{4\}} &=&  \frac{1}{4 \delta (1 +
\epsilon_t^2)^{7/2}} [(4 - \epsilon_t^2) \mu_t^4  + 2(1 +
\epsilon_t^2) (2 - \epsilon_t^2) \mu_t^2 \kappa^2 - \epsilon_t^2
(1 + \epsilon_t^2)^2 \kappa^4 ].
\end{eqnarray}
The pair-production density (per unit spatial volume) from the WKB
actions up through quartic terms is given by
\begin{eqnarray}
{\cal N}^{(0)} \approx  \frac{(q_0E)^{5/2}T}{4 \pi^3 m} (1 +
\epsilon_t^2)^{5/4} e^{- Z_t (\sqrt{1 + \epsilon_t^2} -1)} \Biggl[1
- \frac{5}{16}(4 - 3 \epsilon_t^2) b_t\Biggr]. \label{tem den}
\end{eqnarray}
where $b_t$ is the adiabaticity parameter in Eq. (\ref{bb}). The
pair-production density up to the quadratic terms, the factor in
front of the square bracket, agrees with Eq. (3.40) of Ref.
\cite{DWGS}. We note that Eq. (\ref{tem den}) for the temporal
Sauter-type electric field \cite{Sauter32} can be obtained  by
analytically continuing $\epsilon^2$ to $- \epsilon_t^2$ in Eq.
(\ref{WKB00}) for the Sauter electric field.

\section{Spinor QED}

The eigen-component of the Dirac equation for spin-1/2 fermions
with charge $q$ ($q > 0$) and mass $m$ [in units with $\hbar = c =
1$ and with metric signature $(+, -, -, -)$] takes the form
\cite{Nikishov,Fradkin}
\begin{equation}
 [\eta^{\mu \nu} (\partial_{\mu} + i q A_{\mu}) (\partial_{\nu}
   + i q A_{\nu}) + m^2 + 2 i \sigma qE] \Phi_{\sigma} = 0.
\label{dirac eq}
\end{equation}
The Dirac equation ($\sigma = 1/2$) is the relativistic field
equation for spinor QED, whereas in scalar QED the charged spinless
bosons are described by the Klein-Gordon equation ($\sigma = 0$),
the single component equation in (\ref{dirac eq}) without the
imaginary term.

A comment on the complex instanton actions in spinor QED is in
order. Each eigen-component of the Dirac equation in (\ref{dirac
eq}) satisfies the Klein-Gordon equation with a complex potential
and thus leads to complex instanton actions. For fermions, if we
impose the boundary condition from causality that a particle moves
with a group velocity away from the barrier on the back side, the
expected number of pairs produced per mode is then given by one
minus the reflection probability \cite{Nikishov,Kim-Page2},
\begin{eqnarray}
{\cal N} ({\bf k}_{\perp}) = 1 - \Bigl|\frac{A}{B} \Bigr|^2 = e^{-
(S_{\bf k}+ S^*_{\bf k})} = e^{- ({\cal S}_{{\bf k}_{\perp}} + {\cal
S}^*_{{\bf k}_{\perp}})/2}. \label{fe gr}
\end{eqnarray}
As the imaginary part of an instanton action does {\it not}
contribute to the pair-production rate, from now on we shall refer
to only real part of the instanton action unless stated otherwise.
For a constant electric field, the WKB action in spinor QED is
\begin{eqnarray}
{\cal S}^{(0)}_{sp, {\bf k}_{\perp}} = \frac{\pi(m^2 + {\bf
k}_{\perp}^2)}{qE} + i \pi,
\end{eqnarray}
so  the real part agrees with Eq. (13) of Ref. \cite{Kim-Page} and
Eq. (21) of Ref. \cite{Kim-Page2}, and also with Eq. (27) of Ref.
\cite{Dunne-Schubert} for the action of the worldline instanton.

For the Sauter potential $A_0 (z) = - E_0 L \tanh (z/L)$
\cite{Sauter32}, we change variables to $\zeta = L \tanh (z/L)$ to
write the leading-order WKB instanton action as
\begin{eqnarray}
{\cal S}^{(0)}_{sp, {\bf k}_{\perp}} = 2 S_{{\bf k}_{\perp}} = - i
\oint_{\Gamma_{K}} \frac{\sqrt{-Q_{sp}(\zeta)}}{1 -
\frac{\zeta^2}{L^2}} d\zeta, \label{sp int}
\end{eqnarray}
where
\begin{eqnarray}
Q_{sp}(\zeta) = m^2 + {\bf k}_{\perp}^2 - (\omega + q E_0 \zeta)^2
         + i qE_0 \Bigl(1 - \frac{\zeta^2}{L^2} \Bigr).
\end{eqnarray}
As the contour integral (\ref{sp int}) does not depend on whether
$Q_{sp} (z)$ is real or complex, we can repeat the same procedure as
in scalar QED in Sec. III A. We thus obtain the leading-order WKB
complex action in spinor QED
\begin{eqnarray}
{\cal S}^{(0)}_{sp, {\bf k}_{\perp}} &=&  {\cal S}^{(0)}_{sc, {\bf
k}_{\perp}} + Z \Bigl[ \sqrt{1 + i \pi \delta \epsilon^2} -1 \Bigr],
\label{WKB sp ins}
\end{eqnarray}
where ${\cal S}^{(0)}_{sc, {\bf k}_{\perp}}$ is the WKB action
(\ref{sc ins}) in scalar QED. The real part of the spinor action, which
determines the pair-production rate,
\begin{eqnarray}
{\Re e} \Bigl( {\cal S}^{(0)}_{sp, {\bf k}_{\perp}} \Bigr) = {\cal
S}^{(0)}_{sc, {\bf k}_{\perp}} + \frac{Z}{2} \Bigl[ \sqrt{1 + i \pi
\delta \epsilon^2} + \sqrt{1- i \pi \delta \epsilon^2} - 2 \Bigr]
\label{sp ins}
\end{eqnarray}
is larger than the scalar QED action by $\pi^2 \delta \epsilon^2/4$
up to order ${\cal O} (\delta^3 \epsilon^6)$. We may write the
NLO WKB contribution to the action as
\begin{eqnarray}
{\cal S}^{(2)}_{sp, {\bf k}_{\perp}} = - i
 \oint_{\Gamma_{K}}
\frac{f^{(2)}(\zeta) }{1 - \frac{\zeta^2}{L^2}} d\zeta,
\end{eqnarray}
where
\begin{eqnarray}
f^{(2)} (\zeta) \equiv - \frac{\sqrt{- Q_{sp}}}{32 Q_{sp}^3}
\Biggl[5 \Bigl(\frac{dQ_{sp}}{dz} \Bigr)^2 - 4 Q_{sp} \frac{d^2
Q_{sp}}{dz^2} \Biggr].
\end{eqnarray}
The NLO action is given by
\begin{eqnarray}
{\cal S}^{(2)}_{sp, {\bf k}_{\perp}} = - \frac{\pi^2 \delta
\epsilon^2}{4 \sqrt{1 + i \pi \delta \epsilon^2}},
\end{eqnarray}
so the real part cancels out the excess in the leading-order real
spinor action, at least up to order of ${\cal O} (\delta^3
\epsilon^6)$. Therefore the spinor instanton action up through the
NLO contribution approximately equals the scalar action,
\begin{eqnarray}
{\Re e} \Bigl( {\cal S}^{(0)}_{sp, {\bf k}_{\perp}} + {\cal
S}^{(2)}_{sp, {\bf k}_{\perp}} \Bigr) \approx {\cal S}^{(0)}_{sc,
{\bf k}_{\perp}} \label{sp-sc},
\end{eqnarray}
to the same order. This explains the reason why the leading-order
WKB action in scalar QED gives the pair-production rate closer to
the exact result in spinor QED as explained in Sec. III.

Similarly, for the temporally localized electric field, we obtain
the WKB real action in spinor QED as
\begin{eqnarray}
{\cal S}^{(0)}_{sp, {\bf k}_{\perp}} = {\cal S}^{(0)}_{sc, {\bf
k}_{\perp}} + \frac{Z_t}{2} \Bigl[2 - \sqrt{1 + i \pi \delta
\epsilon^2} - \sqrt{1 - i \pi \delta \epsilon^2}\Bigr].
\end{eqnarray}
The WKB and NLO actions in spinor QED for the temporal electric
field also leads to the WKB action in scalar QED as given by Eq.
(\ref{sp-sc}).

\section{Localized Electric Field and Constant Magnetic Field}

Finally we study the effect of a constant magnetic field $B$, in
the same direction as the electric field, on the production of
charged fermion pairs in the Sauter electric field \cite{Sauter32}
with $\epsilon \ll 1$ and $\epsilon \delta \ll 1$. Note that the
instantons exist when the Landau levels are limited to
\begin{equation}
qB (2 j_{\rm max} + 1) = {\rm min} \{ (\omega + qE_0L)^2 - m^2,
(\omega - qE_0L)^2 - m^2 \}.
\end{equation}
In terms of the potential energy difference
\begin{equation}
V = qA_0(- \infty) - q A_0 (+ \infty) = 2 qE_0 L,
\end{equation}
the highest Landau level is
\begin{equation}
j_{\rm max} = \frac{1}{2qB} \Biggl[\Bigl( \frac{V}{2} - |\omega|
\Bigr)^2 - m^2 \Biggr] - \frac{1}{2}. \label{high lan}
\end{equation}

The pair-production rate per unit area and unit time for spinor QED
is now
\begin{eqnarray}
{\cal N} = \frac{(qB)}{(2 \pi)^2}
 \int_{-(\frac{V}{2} - m)}^{\frac{V}{2} - m} d \omega
\sum_{j = 0}^{j_{\rm max}} \sum_{\sigma_{\pm} = \pm 1/2} e^{ -
{\cal S}_{\sigma_{\pm} j} }, \label{inh fp}
\end{eqnarray}
where $(qB)/(2\pi)$ is the number of Landau levels and another
factor $1/(2\pi)$ is from the $\omega$ integration. Here the
instanton actions are determined by
\begin{equation}
{\cal S}_{\sigma_{\pm} j} = \sum_{n = 0}^{\infty} {\cal
S}^{(2n)}_{\sigma_{\pm} j},
\end{equation}
where the dominant contribution comes from the WKB instantons with action
\begin{eqnarray}
{\cal S}^{(0)}_{\sigma_{\pm} j} &=& -i \oint_{\Gamma_K} dz \sqrt{m^2
+ qB (2j +
1 - 2 \sigma_{\pm}) - (\omega + qE_0 L \tanh(z/L))^2} \nonumber\\
&=& \frac{Z}{2} \Bigl[2 - \sqrt{(1+ \mu)^2 - \epsilon^2 (1 +
\tilde{\kappa}^2) } - \sqrt{(1 - \mu)^2 - \epsilon^2 (1 +
\tilde{\kappa}^2) }\Bigr],
\end{eqnarray}
where $\sigma_{\pm} = \pm 1/2$ and
\begin{eqnarray}
\tilde{\kappa}^2 = \frac{qB (2 j+ 1 - 2 \sigma_{\pm}) }{m^2}.
\end{eqnarray}
Expanding the WKB instanton action up to $\mu^2$ and $\tilde{\kappa}^2$,
summing over $j$ and $\sigma_{\pm}$, and integrating over $\omega$,
we obtain the pair-production rate per unit area and unit time
\begin{eqnarray}
{\cal N}^{(0)} \approx \frac{qBL(qE_0)^{3/2}}{4\pi^2 m} (1 -
\epsilon^2)^{3/4} e^{ - Z (1 - \sqrt{1 - \epsilon^2})} {\rm coth}
\Bigl(\frac{\pi B}{E_0 (1 - \epsilon^2)^{1/2}} \Bigr).
\end{eqnarray}
When $B = 0$, we recover the pair-production rate of Eq.
(\ref{WKB00}), as expected.

\section{Conclusion}

In this paper we have further elaborated the previous instanton
method \cite{Kim-Page,Kim-Page2} by reformulating the instanton
action as a contour integral in the complex plane of space or time.
For a general electric field with the gauge potential $A_0 (z) = -
E_0 f(z)$ or $A_3 (t) = - E_0 f(t)$, $\zeta = f (z)$ or $f(t)$ being
an analytical function, we may express the instanton action as
\begin{eqnarray}
S^{(0)} = - i \oint_{\Gamma_K} \sqrt{(\omega + qE_0 \zeta)^2 - (m^2
+ {\bf k}^2_{\perp})- i qE_0 \Bigl(\frac{df}{dz} \Bigr)} \frac{d
\zeta}{(d f/dz)},
\end{eqnarray}
and a similar formula holds for the time-dependent gauge potential.
As for the Sauter potential, the stratagem is to find the inverse
function $z = f^{-1} (\zeta)$, expand the integrand for large
$\zeta$, and then calculate the contour integral. The inverse
function can be found at least as a power series. This formulation
is gauge independent in the sense that the same instanton action in
the complex plane determines the transmission probability or the
tunneling probability for fermion pair production either in the
time-dependent gauge for time-dependent electric fields or in the
space-dependent gauge for space-dependent electric fields.
Furthermore, the new formulation allows one to calculate the
instanton action beyond the WKB approximation without encountering
the divergence problem at turning points, since it excludes the
branch cut connecting two turning points.

We first applied the instanton method to a constant electric field
both in the space-dependent gauge and in the time-dependent gauge.
In both gauges the instantons recovered the well-known exact result
for the constant electric field. It is shown that the contributions
beyond the leading-order WKB approximation vanish for a constant
electric field, and thus the leading-order WKB approximation is
exact as expected in the one-loop effective action.

We then applied the formulation to a spatially or temporally
localized electric field of Sauter type \cite{Sauter32}. We showed
that our actions agree with the action of the worldline instantons
of Dunne and Schubert \cite{Dunne-Schubert} in the special case of
zero frequency and transverse momentum for a spatially localized
electric field and of zero three-momentum for a temporally
localized electric field. The pair-production rates obtained by
using our leading-order WKB approximations that are expanded up to
quadratic terms of frequency and transverse momentum or
three-momentum also agree with those by the worldline-instanton
method, including the prefactor \cite{DWGS}. Furthermore, the exact
pair-production rate in spinor QED by Nikishov \cite{Nikishov} is
better approximated by the leading-order WKB instanton action in
scalar QED. We show that the cancellation of the leading-order and
the next-to-leading order actions in spinor QED yields the
leading-order scalar QED action and resolves this apparent dilemma.
Finally, we calculate the actions and the pair-production rate in
the Sauter electric field in the presence of a constant magnetic
field.

\acknowledgments

The authors thank G. Dunne and C. Schubert for useful discussions
and F. Khanna for comments on the manuscript. S.P.K. expresses his
appreciation for the warm hospitality of the Theoretical Physics
Institute, University of Alberta. The work of S.P.K. was supported
by the Korea Science and Engineering Foundation under Grant No.
R01-2005-000-10404-0, and the work of D.N.P. by the Natural
Sciences and Engineering Research Council of Canada.

\appendix

\end{document}